\documentclass[11pt]{article}
\usepackage{graphicx}  

\usepackage[T1]{fontenc}
\usepackage[latin1]{inputenc}
\usepackage[all]{xy}
\usepackage{stmaryrd}

\usepackage{xypic}
\usepackage{psfrag}
\usepackage{epsfig}
\usepackage{wrapfig}
\usepackage{a4}
\usepackage{amssymb,latexsym}
\usepackage[mathscr]{eucal}
\usepackage{citesort}
\usepackage{deleq}
\usepackage{url}
\usepackage{fancybox}
\usepackage{ntheorem}

\usepackage{mhequ}

\newcommand{\FS}       
                  {F}

\newcommand{\HS} 
       {H_{\mbox{\scriptsize volume}}}

{\ptc{this should be removed in the oberwolfach version}}%

\newcommand{\eeal}[1]{\label{#1}\end{eqnarray}}
\newcommand{\bed}{\begin{deqarr}}
\newcommand{\eed}{\end{deqarr}}
\newcommand{\bedl}[1]{\begin{deqarr}\label{#1}}
\newcommand{\eedl}[2]{\arrlabel{#1}\label{#2}\end{deqarr}}

\newcommand{\bel}[1]{\begin{equation}\label{#1}}
\newcommand{\bea}{\begin{eqnarray}}
\newcommand{\bean}{\begin{eqnarray}\nonumber}
\newcommand{\beal}[1]{\begin{eqnarray}\label{#1}}
\newcommand{\eea}{\end{eqnarray}}

\newcommand{\Eqsone}[1]{Equations~\eq{#1}}

\newcommand{\qed}{\hfill $\Box$ \medskip}
\newcommand{\proof}{\noindent {\sc Proof:\ }}
\newcommand{\be}{\begin{equation}}
\newcommand{\eeq}{\end{equation}}
\newcommand{\ee}{\end{equation}}
\newcommand{\beqa}{\begin{eqnarray}}
\newcommand{\eeqa}{\end{eqnarray}}
\newcommand{\beqan}{\begin{eqnarray*}}
\newcommand{\eeqan}{\end{eqnarray*}}
\newcommand{\ba}{\begin{array}}
\newcommand{\ea}{\end{array}}

\newcommand{\hyp}{\mycal S}

\newcommand{\mcM}{{\mycal M}}
\newcommand{\mcD}{{\mycal D}}
\newcommand{\dirac}{\mathrm{Dirac}}

\newtheorem{Theorem} {\sc  Theorem\rm} [section]

\newtheorem{Proposition} [Theorem] {\sc  Proposition\rm}

\theorembodyfont{\upshape}

\theoremstyle{nonumberplain}
\theoremsymbol{\ensuremath{\diamondsuit}}
\newcommand{\fcoco}{\small}
\theorembodyfont{\fcoco} \theoremseparator{} \theoremindent0.5cm

\DeclareFontFamily{OT1}{rsfs}{}
\DeclareFontShape{OT1}{rsfs}{m}{n}{ <-7> rsfs5 <7-10> rsfs7 <10-> rsfs10}{}
\DeclareMathAlphabet{\mycal}{OT1}{rsfs}{m}{n}

{\catcode `\@=11 \global\let\AddToReset=\@addtoreset}
\AddToReset{equation}{section}
\renewcommand{\theequation}{\thesection.\arabic{equation}}

\newcounter{mnotecount}[section]

\renewcommand{\themnotecount}{\thesection.\arabic{mnotecount}}

\newcommand{\mnote}[1]
{\protect{\stepcounter{mnotecount}}$^{\mbox{\footnotesize
$
\bullet$\themnotecount}}$ \marginpar{
\raggedright\tiny\em
$\!\!\!\!\!\!\,\bullet$\themnotecount: #1} }

\newcommand{\warn}[1]
{\protect{\stepcounter{mnotecount}}$^{\mbox{\footnotesize
$
\bullet$\themnotecount}}$ \marginpar{
\raggedright\tiny\em
$\!\!\!\!\!\!\,\bullet$\themnotecount: {\bf Warning:} #1} }

\newcommand{\R}{\mathbb R}

\newcommand{\eq}[1]{(\ref{#1})}

\newcommand{\ptc}[1]{\mnote{{\bf ptc:}#1}}

\newcommand{\mcL}{{\mycal L}}
\newcommand{\mcR}{{\mycal R}}

\newcommand{\beqar}{\begin{deqarr}}
\newcommand{\eeqar}{\end{deqarr}}

\newcommand{\beaa}{\begin{eqnarray*}}
\newcommand{\eeaa}{\end{eqnarray*}}

\newcommand{\beq}{\begin{equation}}

\newcommand{\ep}{\epsilon}

\newcommand{\gaz}{\gamma_0}
\newcommand{\Spb}{{\frak S}}

\newcommand{\Rn}{\,{}^{n}\!R}

\newcommand{\tihyp}{\,\,\widetilde{\!\!\hyp}}%
\newcommand{\tcM}{\,\,\,\widetilde{\!\!\!\mcM}}%
\newcommand{\tx}{x}%

\begin{document}

\renewcommand{\dirac}{\mcD}
\title{Killing vectors in asymptotically flat space--times: II.
Asymptotically translational Killing vectors and the rigid
positive
 energy theorem in higher dimensions}

\author{Piotr T.\ Chru\'sciel\thanks{Partially supported by a Polish
Research Committee grant 2 P03B 073 24. E-mail
    \protect\url{Piotr.Chrusciel@lmpt.univ-tours.fr}, URL
    \protect\url{ www.phys.univ-tours.fr/}$\sim$\protect\url{piotr}}
  \\ LMPT, F\'ed\'eration Denis Poisson\\Facult\'e des Sciences\\ Parc
  de Grandmont\\ F37200 Tours, France
  \\
  \\
  Daniel Maerten\thanks{{ E--mail}: maerten@math.univ-montp2.fr} \\
  Institut de Mathématiques et de Mod\'elisation de Montpellier (I3M)\\
  Universit\'e Montpellier II\\
  UMR 5149 CNRS\\
  Place Eug\`ene Bataillon\\
  34095 Montpellier (France)}

\maketitle
\begin{abstract}
We show that the borderline cases in the proof of the positive
energy theorem for initial data sets, on spin manifolds, in
dimensions $n\ge 3$, are only possible for initial data arising
from embeddings in Minkowski space-time.
\end{abstract}

\section{Introduction}
\label{introduction}

Witten's proof of the positive energy theorem~\cite{Witten:mass}
shows that, under appropriate conditions, the time-component of
the energy-momentum vector $p$ is non-negative. For various
reasons it is of interest to understand precisely the borderline
cases, with a vanishing, or perhaps light-like, $p$. In the
context of initial data sets this has been done in detail in an
accompanying paper~\cite{ChBeig1} in space-dimension three. It is
the purpose of this note to generalise the results proved there to
all spin initial data manifolds of dimension $n\ge 3$.

The argument presented in~\cite{ChBeig1} proceeds as follows: in
the borderline cases, Witten's proof provides one or more
covariantly constant ``KIDs'' (by definition, those are the
initial data counterparts of space-time Killing vectors). A
careful study of such KIDs shows that their existence implies the
vanishing of mass, and then flatness of space-time along the
initial data. One then concludes by showing that the Killing
development of the initial data set is flat.

Not unexpectedly, all those arguments can be extended to higher
dimensions, after adjustment of the rates of decay of the fields.
The only part of the proof where essential work is needed is the
algebra proving existence of KIDs. This is based
on~\cite{Maerten}, and presented in Section~\ref{Spet}. On the
other hand, the analysis of the KIDs is essentially identical to
that in~\cite{ChBeig1}, so we will (mainly) only present the
statements of the results needed for the positive energy theorem
here.

The notation and conventions of \cite{ChBeig1} are used throughout. We
assume that the space-dimension $n$ is larger than or equal to three.

Our main results can be summarised as follows:

\begin{Theorem}
  \label{T1new}
  Let $(\mcM,g_{\mu\nu})$ be an $(n+1)$--dimensional space--time, $n\ge 3$, with
a Killing vector field which is  asymptotically null along an
(appropriately regular, see Section~\ref{SKIDs} below)
asymptotically flat spacelike hypersurface $\hyp $.
 Then the  ADM energy--momentum vector of $\hyp $ vanishes.
\end{Theorem}

The precise hypotheses needed for Theorem~\ref{T1new} are the
conditions on the asymptotic behavior of $(g,K)$ in
\eq{V.0}-\eq{V.1} below, together with the matter decay conditions
\eq{V.2} and \eq{K.1501}.
Theorem~\ref{T1new} is a special case of Theorem~\ref{TV.1}
below.

\begin{Theorem}[``Timelike ``future--pointing'' energy--momentum theorem'']
\label{Tpet3} Under natural regularity and matter--energy
conditions (see the conditions of Theorem \ref{Tpet2} below), the
ADM energy--momentum vector $p^ \mu $ of a \underline{spin}
initial-data manifold $\hyp $ satisfies
$$
p^0 > \sqrt{\sum_{i=1}^n(p^i)^2}\;,
$$
unless $(\hyp ,g_{ij},K_{ij})$ are
 initial data for Minkowski space--time.
\end{Theorem}

Theorem~\ref{T1new} is a loose rephrasing of Theorem~\ref{Tpet2}
below.

There are well known counterparts of this with trapped boundaries,
which are of no concern to us here because they always lead to a
strict inequality.

It would be natural to extend the result to cover the Bondi mass, both
in three and higher dimensions. The starting point of the calculations
of the proof of Theorem~\ref{Tpet2} is the existence of a parallel
spinor, the existence of which follows from the analysis in~\cite{CJL}
when the Bondi mass is null in space-dimension three. The calculations
that follow apply without modifications, yielding a parallel isotropic
KID. One expects that this is incompatible with a non-vanishing
Trautman-Bondi mass, but a complete analysis of this has not been
carried out so far.



\section{KIDs in $n$-dimensional asymptotically flat
initial data sets, $n\ge 3$} \label{SKIDs}

We have the following string of propositions, which are the
building stones of the proof of Theorem~\ref{TV.1} below:

\begin{Proposition}\label{2PN.1} Let $R > 0$ and let $(g_{ij},K_{ij})$ be
initial data on $\hyp _R \equiv \R^n \setminus B(R)$ satisfying \beq
g_{ij} - \delta_{ij} = O_{k} (r^{-\alpha}), \qquad
 K_{ij} = O_{k-1} (r^{-1-\alpha}),
\label{K.99}
\eeq
with some $ k>1$ and some $\alpha>0$. Let $N$ be a $C^2$ scalar field
and $Y^i$ a $C^2$ vector field on $\hyp _R$ such that
\beq \label{K.4}
2NK_{ij} + {\cal L}_Y g_{ij} = 0 \;.
\eeq
Define $\rho$, $J^i$ and $\tau_{ij}$ by the equations
\begin{eqnarray}
\label{K.13} &
 2 \rho = {}^nR+(K^i{_i})^2-K^{ij}K_{ij}\ ,
& \\ & J^i=D_j(K^{ij}-K^k{_k}g^{ij}) \ , & \label{K.14} \\ &
\tau_{ij}-\frac{1}{2}g^{k\ell}\tau_{k\ell}g_{ij} = {}^nR_{ij} +
K^k{_k}
K_{ij} - 2 K_{ik} K^k{}_j \qquad\qquad\qquad\qquad \nonumber & \\
& \qquad \qquad \qquad\qquad\qquad - N^{-1}(\mcL_Y K_{ij} + D_i
D_j N  ) -\frac{\rho}{2}\,g_{ij} \ , & \label{K.15}
\end{eqnarray}
and assume that $\rho$ and $\tau_{ij}$ satisfy
\begin{equation}
\rho= O_{k-2} (r^{-2-\alpha})\;, \qquad \tau_{ij}= O_{k-2}
(r^{-2-\alpha})\;.
\label{K.100}
\end{equation}
  Then there exists numbers $ \Lambda_{\mu\nu}= \Lambda_{[\mu\nu]} $
  such that we have, for $r$ large,
\beal{D.4.0.a} & D_iY_j-\Lambda_{ij} = O_{k-1}(r^{-\alpha})\;,
\qquad
Y^i-\Lambda_{ij} x^j = \left\{%
\begin{array}{ll}
     O(r^{1-\alpha}), & \hbox{$\alpha\ne 1$;} \\
    O(\ln r), & \hbox{$\alpha=1$,} \\
\end{array}%
\right.
 &
 \\
 &
D_iN-\Lambda_{i0} = O_{k-1}(r^{-\alpha})\;,
\qquad N -\Lambda_{i0}x^i = \left\{%
\begin{array}{ll}
     O(r^{1-\alpha}), & \hbox{$\alpha\ne 1$;} \\
    O(\ln r), & \hbox{$\alpha=1$.} \\
\end{array}%
 \right. &\label{D.4.0} \eea
If ${\Lambda_{\mu\nu}}=0$, then there exist numbers $A^\mu$
 such that we have
\begin{equation}
  Y^i-A^i  = O_{k}(r^{-\alpha}), \qquad
N-A^0 = O_{k}(r^{-\alpha}) \ .
  \label{D.4.1}
\end{equation}
If ${\Lambda_{\mu\nu}}=A^\mu=0$, then  $Y^i\equiv N \equiv 0$.
\end{Proposition}

\proof See Section~2 and Appendix~C of~\cite{ChBeig1}. \qed

\begin{Proposition}\label{PN.1}
 Let $R > 0$ and let $(g_{ij},K_{ij})$ be
initial data on $\hyp _R $ satisfying
\begin{eqnarray} \label{F0.1} &
g_{ij} - \delta_{ij} = O_{2} (r^{-\alpha}), \qquad
 K_{ij} = O_{1} (r^{-1-\alpha}),
 \qquad \alpha > (n-2)/2, & \\
& \label{F0.2}
 J^i = O (r^{-n-\epsilon}), \qquad \rho = O (r^{-n-\epsilon}),
\qquad \epsilon > 0\ . &
\end{eqnarray}
Let $N$ be a $ C^1$ scalar field and $Y^i$  a $C^1$ vector field on
$\hyp _R$ such that \beq \label{F0.1.1} N-A^0 = O_{1} (r^{-\alpha}),
\qquad Y^i\to _{r\to\infty} A^i \ , \eeq
 for some set of constants $(A^\mu) \not\equiv 0$, satisfying
\begin{equation}
2 N  K_{ij} + \mcL_Y g_{ij} =  O_1 (r^{-(n-1)-\epsilon}).
\label{(PN.1.0)}
\end{equation}
Let $p^\mu$ be the ADM energy--momentum  of\/  $\hyp _R$. Then:
\begin{enumerate}
\item If $A^0 = 0$, then $p^0 = 0$.
\item  If $A^0 \ne 0$, then  $p^\mu$ is proportional to
$A^\mu$.
\end{enumerate}
\end{Proposition}

\proof See the proof of Proposition~3.1 in~\cite{ChBeig1}. \qed

\begin{Proposition}\label{PN.2}
Under the hypotheses of Proposition \ref{PN.1}, suppose further that
$N$ is $C^2$ and that \beq
 \tau_{ij} =  O (r^{-n-\epsilon}) \ .  \label{F0.1.12} \eeq
If \beq \label{F0.1.21} (A^0)^2 < \sum_i A^iA^i\ , \eeq then $p^\mu$
vanishes.
\end{Proposition}

\proof See the proof of Proposition~3.2 in~\cite{ChBeig1}. \qed

\begin{Proposition}\label{PN.1.1} Under the hypotheses of Proposition
\ref{PN.1}, assume moreover that $N$ is $C^2$, that
 \eq{F0.1.12} holds and that
\begin{eqnarray}  \label{X.0.1} &
 N K_{ij}+ D_iY_j = O_1(r^{-(n-1)-\epsilon}) \ , &
\\ \label{X.0.2}
&  K_{ij}Y^j+ D_i N = O_1(r^{-(n-1)-\epsilon})  \ ,  &
\\  &
A^\mu A_\mu \neq 0.  &\nonumber
\end{eqnarray}
Then the ADM energy--momentum $p^\mu$ vanishes.
\end{Proposition}

\proof See the proof of Proposition~3.3 in~\cite{ChBeig1}.  Note
that the proof in~\cite{ChBeig1} uses the equality of the Komar
and the ADM masses for translational, asymptotically timelike
Killing vectors, while Proposition~\ref{PN.2} shows that one only
needs to consider timelike $A^\mu$'s to complete the proof.  The
equality of those masses, which is well known in space-dimension
three~\cite{BeigKomar}, can also be established in higher
dimensions by an asymptotic analysis of the stationary Einstein
equations when the sources decay sufficiently fast. \qed

The notation used in the next theorem is explained in
Appendix~\ref{asymptotic}:

\begin{Theorem}\label{TV.1}
Let $R>0$ and let $(g_{ij},K_{ij})$ be initial data on $\hyp
_R=\R^n\backslash B(R)$ satisfying
\begin{eqnarray} &
g_{ij}-\delta_{ij}=O_{3+\lambda}(r^{-\alpha}), \qquad
K_{ij}=O_{2+\lambda}(r^{-1-\alpha}), & \label{V.0}  \\
&\alpha> \left\{
           \begin{array}{ll}
             1/2, & \hbox{$n=3$;} \\
             n-3 , & \hbox{$n\ge 4$,}
           \end{array}
         \right.
 \qquad \epsilon >0, \qquad  0<\lambda<1. & \label{V.1} \\
& J^i = O _{1+\lambda}(r^{-n-\epsilon}), \qquad \rho =
O_{1+\lambda} (r^{-n-\epsilon})\;.\label{V.2}
\end{eqnarray}
Let $N$ be a scalar field and $Y^i$  a vector field on $\hyp _R$
such that
$$
N\to_{r\to\infty}A^0,\quad Y^i\to_{r\to\infty}A^i, \qquad A^\mu
A_\mu=0\ ,
$$
for some constants $A^\mu\not \equiv 0$. Suppose further that
\begin{eqnarray} &
2 N  K_{ij} + \mcL_Y g_{ij} =  O_{3+\lambda} (r^{-(n-1)-\epsilon})\
, & \label{K.1500}
\\ &
\tau_{ij} =O_{1+\lambda} (r^{-n-\epsilon})\;,  & \label{K.1501}
\end{eqnarray}
Then the ADM energy--momentum of $\hyp  _R$ vanishes.
\end{Theorem}

\proof See the proof of Theorem~3.4 in~\cite{ChBeig1}. We note
that in our context \cite[Equation~(3.40)]{ChBeig1} reads
 \begin{equation}
g_{nA}=\left\{
         \begin{array}{ll}
          C_{AB}(x^n)\partial_B \ln\rho +O_{(1)}( \rho ^{-1-\ep}\ln \rho), & \hbox{$n=3$;} \\
           C_{AB}(x^n)\partial_B \frac 1 {\rho^{n-3}} +O_{(1)}( \rho ^{-(n-2)-\ep}), & \hbox{$n\ge 4$.}
         \end{array}
       \right.
 \label{V.2.2}
 \end{equation}
 Similarly instead of
 \cite[Equation~(3.47)]{ChBeig1} we have
  \begin{equation}
\frac{\partial g_{AB}}{\partial x^n}= \left\{
                                        \begin{array}{ll}
                                    D_{ABCD} \partial_C\partial_D \ln \rho+ O_{(1)}( \rho ^{-2-\ep}\ln
\rho)\;, & \hbox{$n=3$;} \\
                                   D_{ABCD} \partial_C\partial_D \frac 1 {\rho^{n-3}}+
O_{(1)}( \rho ^{-(n-1)-\ep})\;, & \hbox{$n\ge 4$.}
                                        \end{array}
                                      \right.
\label{V.4}
 \end{equation}
Finally, there are  misprints in the definitions of the quantities
$\Omega$ and $\Omega'$ in the proof there; the correct
definitions, in all dimensions, are\footnote{We take this
opportunity to point out that equation (2.20) of \cite{ChBeig1}
(which is equation (2.27) of the gr-qc version of that paper)
should be replaced by $\rho= O_{k-2} (r^{-2-\alpha})$, $
\tau_{ij}= O_{k-2} (r^{-2-\alpha})$. Furthermore, Equations (2.15)
and (3.34) of \cite{ChBeig1} are mutually incompatible; the
correct one is (2.15).}
$$
\Omega=\lim_{\rho \to\infty}\sum_C \int_{S^{n-2}(\rho ,x^n)}
(x^A\partial_C g_{nA}-g_{nC})dS_C\;.
 $$
 $$
\Omega'=\lim_{\rho \to\infty} \int_{S^{n-2}(\rho ,x^n)}
\Big((n-1)(x^A x^B
\partial_C\partial_n g_{AB}-2x^B\partial_n g_{CB})-x^A x^A\partial_C\partial_n g_{BB}+ 2
x_C \partial_n g_{AB}\Big) dS_C\;,
$$
where summation over every repeated occurrence of indices is
implicitly understood, regardless of their positions. Here
$\rho^2=(x^1)^2+\ldots+(x^{n-1})^2$, while $S^{n-2}(\rho,a )$ is a
sphere (or circle, when $n=3$) of radius $\rho $ centred at
$x^1=\ldots=x^{n-1}=0$ lying in the plane $x^n=a$. Finally the
$dS_C$'s are the usual surface forms $dS_C=\partial_C\rfloor (
dx^1\wedge \cdots \wedge dx^{n-1})$, and $\rfloor$ denotes
contraction. \qed

\section{The rigid positive energy theorem}\label{Spet}

The following strengthens somewhat Theorem~4.1 of~\cite{ChBeig1}
in the case $n=3$, and generalises that theorem to higher
dimensions; the calculations here are closely related to those
in~\cite{Maerten}:

\begin{Theorem}[(Rigid) positive energy theorem]
\label{Tpet}
 Consider a data set
 $(\hyp ,g_{ij},K_{ij})$, with $(\hyp,g_{ij}) $ a complete
Riemannian spin manifold of dimension $n\ge 3$, and with
$g_{ij}\in C^2$, $K_{ij}\in C^1$. Suppose that $\hyp$ contains an
asymptotically flat end $\hyp _R$ diffeomorphic to
${\R}^n\setminus B(R)$ for some $R>0$, with $B(R)$ --- a
coordinate ball of radius $R$, where the fields $(g,K)$ satisfy
\begin{equation}
\label{falloff} | g _{ij}-\delta_{ij}|+|r\partial_k  g
_{ij}|+|rK_{ij}|  \le C r^{-\alpha}\ ,
\end{equation}
for some  constants $C>0$ and $\alpha>\min(1/2,n-3)$, with
$r=\sqrt{\sum_{i=1}^n (x^i)^2}$.
  Suppose moreover that the quantities $\rho$ and $J$
\begin{eqnarray}
& 2 \rho := {}^3R + (K^k{_k})^2 - K^{ij}K_{ij}\,, &
  \label{EP.1}\\
& J^k := D_l (K^{kl} - K^k{_k} g^{kl}) \,, &
  \label{EP.2}
\end{eqnarray}
satisfy
  \begin{equation}
\sqrt{ g _{ij}J^iJ^j}\le \rho \le C(1+r)^{-n-\epsilon}, \qquad
\epsilon >0.
    \label{EP.3}
  \end{equation}
Then the
 ADM energy--momentum $(m,p^i)$ of any of the asymptotic ends of $\hyp $
satisfies \bel{posmasthm}m\ge \sqrt{p_ip^i}\;.\ee If $m=0$, then  $\rho\equiv J^i
\equiv 0$, and there exists an isometric embedding $i$ of $\hyp $
into Minkowski space--time $({\R}^{n+1},\eta_{\mu\nu})$ such that
$K_{ij}$ represents the extrinsic curvature tensor of $i(\hyp )$
in $(M,\eta_{\mu\nu})$. Moreover $i(\hyp )$ is an asymptotically
flat Cauchy surface in $({\R}^{n+1},\eta_{\mu\nu})$.
\end{Theorem}

 Theorem~\ref{Tpet2} has been formulated under
differentiability requirements which are stronger than necessary,
compare~\cite{BartnikChrusciel1,Miaomath-ph/0212025}.
Unfortunately our proof that ADM energy--momentum cannot be null
requires even more differentiability and asymptotic decay
conditions:

\begin{Theorem}
\label{Tpet2} Under the hypotheses of Theorem \ref{Tpet}, suppose
moreover that
\begin{eqnarray} &
g_{ij}-\delta_{ij}=O_{3+\lambda}(r^{-\alpha}), \qquad
K_{ij}=O_{2+\lambda}(r^{-1-\alpha}), & \label{V.00}  \\
& \rho=O_{1+\lambda}(r^{-n-\ep}), & \label{V.01}
\end{eqnarray}
with some $0<\lambda<1$. Then the ADM energy--momentum cannot be
null.
\end{Theorem}

{\noindent \sc Proofs of Theorems~\ref{Tpet} and \ref{Tpet2}:}
 We  use a Witten-type argument, as
follows. Let $(\Spb,\langle\cdot,\cdot\rangle)$ be any Riemannian
bundle of real spinors over $(M,g)$ with scalar product
$\langle\cdot,\cdot\rangle$, such that Clifford multiplication
(which we denote by $X\cdot$) is anti-symmetric. We suppose that
there exists a bundle isomorphism $\gaz:\Spb\to\Spb$ with the
following properties: \minilab{prop1}\begin{equs}\label{prop1a}  &
\gaz ^2 = 1\;,
 \\
 \label{prop1b}
 &\forall X\in TM\quad  \gaz X \,\cdot= -X \cdot \gaz
  \;,
 \\
 \label{prop1c}
 & ^t \gaz = \gaz
  \;,
  \\
 \label{prop1d}
 & D\gaz = \gaz D
  \;,
 \end{equs}
where $^t\gaz$ denotes the transpose of $\gaz$ with respect to
$\langle\cdot,\cdot\rangle$, and $D$ is the usual Riemannian
spinorial connection associated with the metric $g$.

(Such a map always exists if $\Spb$ is obtained by pulling-back a
space-time spinor bundle, using an  externally oriented isometric
embedding of $(M,g)$  in a Lorentzian space-time. Then the
Clifford product $n\,\cdot$, where $n$ is the field of Lorentzian
unit normals to the image of $M$, has the required properties. If,
however, such a map does not exist, we proceed as follows: let
$\Spb'=\Spb\oplus\Spb$ be the direct sum of two copies of $\Spb$,
equipped with the direct sum metric
$\langle\cdot,\cdot\rangle_\oplus$:
\bel{prop2.0}\langle(\psi_1,\psi_2),(\varphi_1,\varphi_2)\rangle_\oplus :=
\langle\psi_1,\varphi_1\rangle+
 \langle\psi_2,\varphi_2\rangle\;.\ee
We set, for $X\in TM$, \minilab{prop2}\begin{equs}\label{prop2a} &
\gaz (\psi_1,\psi_2) := (\psi_2,\psi_1)\;,
 \\
 \label{prop2b}
 &
    X \cdot (\psi_1,\psi_2):= (X\cdot \psi_1,-X\cdot \psi_2)\;,
 \\
 \label{prop2c}
 &
   D_X (\psi_1,\psi_2):= (D_X\psi_1,D_X \psi_2)
  \;.
 \end{equs}
One readily verifies that \eq{prop2b} defines a representation of
the Clifford algebra on $\Spb'$, and that \eq{prop1} holds.)

Given an initial data set $(M,g,K)$,  a vector field $X$, and a
spinor field $\xi$ we set
 \beal{Kprddef}
 K(X) &:=& K_{i}{^j}X^i e_j\,\cdot\;,
 \\
\nabla_{X}\xi & := & D_{X}\xi + \frac{1}{2}K(X)\gaz  \xi\;.
 \eeal{nezcondef} Here $e_i$ is a local orthonormal basis of $TM$; it
 is straightforward to check that \eq{Kprddef} does not depend upon
 the choice of this basis. (To make things clear, \eq{nezcondef}
 defines $\nabla$ in terms of the Riemmanian spin connection $D$. If
 the spin bundle arises from a space-time bundle, then
 $\nabla$ coincides with the canonical space-time spinorial
 derivative, when restricted to space directions.)

We will need an explicit expression for the curvature of $\nabla$:

\begin{Proposition}\label{PSpcurvature} {\it For every $X ,Y \in \Gamma(T\hyp)$ we have
\bel{ePSpcu} R_{X,Y}=  {\Rn_{X,Y}} +\frac{1}{2}\text{d}^{D}K(X,Y)\gaz   -
\frac{1}{4}\Big(K(X) K(Y)-K(Y) K(X)\Big)
 \;,\ee
where $R$ is the curvature of $\nabla$, $\Rn$  is that of $D$,}
and
$$\text{d}^{D}K(e_i,e_j)= (K^k{}_{j;i}-K^k{}_{i;j})e_k\,\cdot\;.$$
\end{Proposition}

\proof We have
\begin{eqnarray*}
\nabla_{X}\nabla _{Y} \psi & = & \Big(D_{X}
+\frac{1}{2}K(X)\gaz\Big) \Big( D_{Y}\psi +\frac{1}{2}K(Y)\gaz \psi
 \Big)
 \\
 &=&
 D_{X}D_{Y} \psi  +\frac{1}{2}K(X)\gaz  D_{Y}\psi
 \\
 &&+\frac{1}{2}\Big( (D_{X}K)(Y)\gaz \psi  +K(D_{X}Y)\gaz \psi  +K(Y)\gaz  D_{X}\psi \Big)
 \\
 &&
 + \frac 14 K(X)\gaz K(Y)\gaz \psi
 \\
 &=&  D_{X}D_{Y} \psi  +\frac{1}{2}\Big(K(X)\gaz  D_{Y}\psi+K(Y)\gaz
 D_{X}\psi\Big)
 \\
 && +\frac{1}{2}\Big((D_{X}K)(Y)\gaz \psi +K(D_{X}Y)\gaz \psi  \Big)
 \\
 &&-\frac{1}{4}K(X) K(Y) \psi
 \;,
\end{eqnarray*}
so that \beaa R_{X,Y}\psi &=& \nabla_X \nabla_Y \psi - \nabla_Y
\nabla_X \psi -\nabla_{[X,Y]} \psi
 \\
 &=& D_X D_Y \psi - D_Y
D_X \psi -D_{[X,Y]}\psi - \frac 12 K([X,Y]) \gaz \psi
 \\
 && +\frac{1}{2}\Big(((D_{X}K)(Y)-(D_{Y}K)(X))\gaz \psi +K(D_{X}Y-D_{Y}X)\gaz \psi  \Big)
 \\
 &&-\frac{1}{4}\Big(K(X) K(Y)- K(Y)K(X)\Big) \psi
 \;,
\end{eqnarray*}
and the vanishing of the torsion of the Levi-Civita connection gives
the result.

\qed

We now run the usual Witten argument (see,
\emph{e.g.},~\cite{BartnikChrusciel1}) using the connection $\nabla$
and the associated Dirac operator $\dirac=e^i\cdot\nabla_i$. Under the
current conditions the ADM energy--momentum of $\hyp $ is finite and
well defined \cite{ChErice,Bartnik86}, and the Witten boundary
integral reproduces the ADM energy--momentum. The arguments in
\cite{BartnikChrusciel1} show that, again under the current
conditions, for every spinor field $\mathring \psi$, with constant
entries in the natural spin frame in the asymptotic region, one can
find a solution $\psi$ to the Witten equation which asymptotes to
$\mathring \psi$. Witten's identity subsequently implies that
\bel{posCla}
\langle \mathring \psi, p \cdot \mathring \psi \rangle \ge 0\;, 
\ee
where $$p\;\cdot:=m \gaz +p^ie_i\;\cdot\;,$$  and $p=(m,p^i)$ is
the ADM momentum. This gives \eq{posmasthm}.

The equality case, which is
of main interest here, is only possible if $p$ is lightlike or
vanishes. In either case one obtains a spinor field $\psi \in
\Gamma(\Spb)$ which is asymptotic to $\mathring{\psi}$, and
satisfies
\bel{paral} \nabla \psi = 0\;,\ee \bel{rigid} \langle \psi,\mcR
\psi\rangle = 0\;.\ee Here
 $$\mcR :=\frac{1}{2} \left(\rho +J^ie_i \cdot\gaz  \right)$$ is
the (non-negative) spinorial endomorphism which appears in the
identity:
$$\dirac ^{*}\dirac =\nabla^{*}\nabla +\mcR \;.$$

 The idea of the calculations that follow is to show, roughly
speaking, that the space-time is a pp-wave space-time, perhaps with
matter decaying at inifinity, with a null Killing vector, which by the
results in the previous section is only possible if we are in
Minkowski space-time. We start with an analysis of the curvature
tensor.

  As $\psi $ is $\nabla $-parallel we have $R_{XY}\psi =0$, and from
Proposition~\ref{PSpcurvature} one finds, for all $X,Y\in T\hyp$,
$$\langle {\Rn _{X,Y}}\psi,\psi\rangle
-\frac{1}{2}\langle\text{d}^{D}K(X,Y)\cdot \gaz \psi,\psi \rangle
- \frac{1}{4}  \langle \big(K(X) K(Y)-K(Y) K(X)\big)\psi,\psi \rangle=0.$$
Both the first and third term vanish since the spinorial curvature
can be written as
$$\Rn _{XY}\psi = - \frac{1}{2}  \sum_{i<j} \ \Rn (X,Y,e_{i},e_{j}) e_{i}\cdot e_{j} \cdot \psi \;,$$
and since the Clifford product of two distinct elements of an ON
basis is anti-symmetric. (We use the  conventions
 $$\Rn(e_i,e_j)e_k=
D_{e_i}D_{e_j}e_k-D_{e_j}D_{e_i}e_k
-D_{[e_i,e_j]}e_k=\Rn^s{}_{kij}e_s=\Rn(e_m,e_k,e_i,e_j)g^{sm}e_s\;,$$
 $$\Rn_{ij}=\Rn^k{}_{ikj}\;,$$
where ${\Rn}_{ij}$ is the Ricci tensor of $g$.)
Thus we obtain
\bel{DerKc}
\langle\text{d}^{D}K(X,Y) \gaz \psi,\psi \rangle=0
\;.
\ee
Let us denote by $N$ the function
\bel{Ndef}
N=\langle\psi,\psi\rangle\;,
\ee
 and by $Y$ the real 1-form defined as
\bel{Ydef}
Y(X)= -\langle \gaz X \cdot \psi,\psi \rangle\;.
\ee
%
Using
this notation, \eq{DerKc} can be rewritten as
\bel{DerKc2}
K_{ki;j}Y^k=K_{kj;i}Y^k\;.
\ee
%
%
We continue with the following calculation:
 \bean \sum_{k=1}^{n} e_{k} \cdot R_{e_{s},e_{k}}
&=& \sum_{k=1}^{n} e_{k} \cdot \Big(\Rn _{e_{s},e_{k}}
  - \frac{1}{4}\Big(K(e_s)
K(e_k)-K(e_k) K(e_s)\Big)
 \\
 \nonumber
 &&+\frac{1}{2}\text{d}^{D}K(e_{s},e_{k})\gaz\Big)
 \\
  &=& - \frac{1}{4} \Big(\Rn _{s}{^{kij}}+ K_s{^i} K^{kj}
-K^{ki} K_s{^j} \Big)e_{k} \cdot  e_i\cdot e_j\cdot
\nonumber
 \\
 &&+\frac{1}{2}(K^{mk}{}_{;s}-K^m{}_{s}{}^{;k})e_{k} \cdot e_m\cdot
 \gaz
\;.
 \eeal{ContrRiem}
 In order to analyse the curvature terms in the before-last line of
\eq{ContrRiem}, recall the convenient identity\footnote{To prove
\eq{convid}, note first that the result is clearly true if all indices
are distinct or equal; the final formula follows by inspection of the
remaining possibilities.}
 \bel{convid} e_k\cdot e_i \cdot e_j \cdot =
e_{[k}\cdot e_i \cdot e_{j]} \cdot - g_{ki} e_j\cdot + g_{ij} e_k
\cdot - g_{kj} e_i\cdot \;.
\ee
(Square brackets around indices denote anti-symmetrisation, and round
brackets denote symmetrisation.) The Bianchi identity $\Rn
_{s}{^{[kij]}}=0$ immmediately implies
%
\[
 \Rn _{s}{^{kij}} e_k\cdot e_i \cdot e_j \,\,\cdot= 2 \;\Rn _{s}{^{i}}e_i\,\cdot\;.
 \]
 Next, the undifferentiated extrinsic curvature terms in 
next-to-last line of \eq{ContrRiem} can be manipulated as
\beaa \lefteqn{K_s{^i} K^{kj} \underbrace{e_{k} \cdot  e_i
 \,\cdot}_{-2g_{ki}-e_i \cdot e_k\,\cdot} e_j\cdot
-K_s{^j} \underbrace{K^{ki} e_{k} \cdot
e_i\,\cdot}_{-K^{ki}g_{ki}} e_j\,\cdot}
 &&
 \\
 &&=-2 K_s{^i} K^{kj} g_{ki}e_j\cdot- K_s{^i}e_{i} \cdot \underbrace{K^{kj}  e_k
 \cdot e_j\,\cdot}_{-K^{kj}g_{kj}}
+K_s{^j} K^{ki} g_{ki} e_j\cdot
 \\
 &&=
2\Big(-   K^{kj} K_{sk}+K^k{_k}K{_{s}{}^j}\Big)
 e_j\cdot\;,
 \eeaa
  which results in
  \bean \lefteqn{\Big(\Rn
_{s}{^{kij}}+K_s{^{i}} K^{kj} -K^{ki} K_s{^{j}}\Big) e_k\cdot e_i
\cdot e_j \,\cdot}&&
 \\
 && =2\Big( \Rn _{s}{^{i}}+ K^k{_k}K{_{s}{}^i}-   K^{ki}
K_{sk}\Big)
 e_i\,\cdot=: 2 E_s{^i}e_i \,\cdot=: 2E(e_s)\;.
 \eeal{finseclin}
Using again that $\psi $ is $\nabla $--parallel we have
$\sum_{k=1}^{n} e_{k} \cdot R_{e_{s},e_{k}}\psi =0$.  \Eqsone{ContrRiem} and \eq{finseclin}
show that
  $$\left(E(e_s) -
(K^{mk}{}_{;s}-K^m{}_{s}{}^{;k})e_{k} \cdot e_m\cdot\gaz
\right)
\psi=0
\;. $$
Multiplying by $e_r\cdot$ and taking a scalar product with $\psi$ we obtain
\bean
-N E_{rs}
&=&
 (K^{mk}{}_{;s}-K^m{}_{s}{}^{;k}) \langle \psi,
\underbrace{e_r \cdot e_{k} \cdot e_m\cdot}_{=e_{[r}\cdot e_k\cdot e_{m]}\cdot -g_{rk}e_m\cdot
+g_{rm}e_k\cdot - g_{km}e_r\cdot} \gaz
\psi\rangle
\\
& = &
 (K^{mk}{}_{;s}-K^m{}_{s}{}^{;k}) \langle \psi, (-g_{rk}e_m\cdot
+g_{rm}e_k\cdot - g_{km}e_r\cdot)
\gaz
\psi\rangle
\nonumber \\
& = &
 (K^{mk}{}_{;s}-K^m{}_{s}{}^{;k}) (-g_{rk}Y_m+g_{rm}Y_k- g_{km}Y_r)
\nonumber \\
& = &
 -(K_{rs;k}-K_{ks;r})Y^k+J_sY_r \;,
\eeal{tod3}
where we have used the fact that the products $e_r\cdot e_m$ and $e_r
\cdot e_{k} \cdot e_m\cdot \gaz$ are anti-symmetric when all indices
are distinct, and therefore give no contribution in \eq{tod3}.
Hence
\bel{RicKvan} N\left({\Rn}_{ij}+K^{k}{}_{k}K_{ij} - K_{ik}
K^k{}_{j}\right)= (K_{ij;k}-K_{kj;i})Y^k-J_jY_i\;.\ee
Taking a trace implies
\bel{rhovan} N\rho=-J^iY_i\;.
\ee
Anti-symmetrising \eq{RicKvan} in $i$ and $j$ and using \eq{DerKc2} one finds
\bel{tod4}
J_i=\sigma Y_i
\ee
for some function $\sigma$.

We wish, now, to show that the pair $(N,Y^{i})$ defined by
\eq{Ndef}-\eq{Ydef} satisfies \eq{K.4}. It is convenient to choose an
ON basis $\{e_{i}\}^{n}_{i=1}$ which satisfies $e_{i}=\partial_i$ and
$D_{e_{i}}e_{j}=0$ at the point under consideration, then
\begin{eqnarray*}
-D_{i}Y_{j} &=& \partial_{i}\langle \gaz  e_{j}\cdot \psi ,\psi
\rangle
 =
 \langle \gaz  e_{j}\cdot D_i \psi ,\psi  \rangle
 +\langle \gaz  e_{j}\cdot \psi ,D_i \psi  \rangle
  \\
 &=&
 2 \langle \gaz  e_{j}\cdot
 \underbrace{D_i \psi}_{-\frac 12 K_i{^k}e_k\cdot\gaz \psi} ,\psi  \rangle
 =- K_i{^k}\langle  e_{j}\cdot
 e_k\cdot \psi ,\psi  \rangle
 \\
 &=&
 - K_i{^k}\underbrace{\langle  e_{[j}\cdot
 e_{k]}\cdot \psi ,\psi  \rangle}_{0}
 - K_i{^k}\langle  \underbrace{e_{(j}\cdot
 e_{k)}\cdot}_{-g_{jk}} \psi ,\psi  \rangle
 \\
&=& N K_{ij}\;,
\end{eqnarray*}
as desired. 

Next,
\begin{eqnarray*}
D_{i}N
 &=& \partial_{i} \langle \psi ,\psi  \rangle
 = 2 \langle \psi ,D_i \psi  \rangle
 = - \langle \psi ,K_i{^k}e_k\gaz \psi  \rangle
 \\
&=& -K_{ik}Y^{k} 
\end{eqnarray*}
(compare \eq{X.0.2}). For further use we note that
$d(N^2-|Y|^2)=0$, and as $N^2-|Y|^2\to_{r\to \infty}0$ (since equality
is attained in \eq{posCla}) we conclude that
$$N^2=|Y|^2\;.$$
Further differentiation yields
 $$
 D_{i}D_{j}N= N(K\circ
K)_{ij}-D_iK_{jk}Y^{k}\;.
$$
Inserting this into \eq{K.15} and
using the relations above
leads to our key formula
\bel{tauvanfu} N^2\tau_{ij} =\rho Y_i Y_j\;.
\ee
%
Note that $N\to_{r\to\infty}0$ implies $Y\to _{r\to\infty}0$. The last
part of Proposition~\ref{2PN.1} gives then $N\equiv 0$, hence
$\psi=0$, contradicting the fact that we have a non-trivial solution
of the Witten equation. Thus $N$ approaches a non-zero constant at
infinity by \eq{D.4.1}, and our hypothesis on the decay of $\rho$
provides decay of $\tau_{ij}$. We can therefore apply
Proposition~\ref{PN.1.1} and Theorem~\ref{TV.1} to conclude that the
ADM momentum vanishes. But then for any
$\mathring \psi$ there exists an associated $\nabla$-parallel
$\psi$. Let $\mathring\psi_a$, $a=1,\ldots, m$, form a basis and let
$\psi_a$ be the parallel spinor that asymptotes to
$\mathring\psi_a$. Now,
$$
\nabla \langle \psi_a,\psi_b\rangle = 0\;,
$$
 which implies that
the $\psi_a$'s form a basis of $\Spb_p$ at every $p\in \hyp$. It
follows that $R_{XY}\psi_a=0$ for a collection of spinors forming
a basis at each point, hence
\bel{curvvanfu} R_{XY}=0\;.\ee

Choose $\mathring \psi$ so that $N\to 1$ and $Y\to 0$. (If no such
$\mathring \psi$ exists, we pass to $\Spb'$ with the structures
defined by \eq{prop2.0}-\eq{prop2}, choose any $\mathring \chi$
with norm one-half, then $\mathring \psi = (\mathring \chi,
\mathring \chi)$ will have the desired property.) Let $\tihyp$ be
the universal
 covering space of $\hyp$ with corresponding data $\left(\tihyp
,\widetilde{g}_{ij},\widetilde{K}_{ij},\widetilde{N},\widetilde{Y}^{j}
\right)$, and consider  the \emph{Killing development} thereof:
 by definition, this  is  $\tcM = \mathbb{R}\times \tihyp $
endowed with the metric
$$ \widetilde{g}_{\mu\nu}=
-\widetilde{N}^{2}\text{d}u^{2}+
\widetilde{g}_{ij}\left(\text{d}x_{i}+ \widetilde{Y}^{i}\text{d}u
\right)\left(\text{d}x_{j}+ \widetilde{Y}^{j}\text{d}u   \right)
\;,
$$ 
where $\widetilde{N}(u,\tx )=\widetilde{N}(\tx ), \quad
\widetilde{g}_{ij}(u,\tx )= \widetilde{g}_{ij}(\tx ), \quad
\widetilde{Y}^{j}(u,\tx )= \widetilde{Y}^{j}(\tx )$. Similarly let
$(\mcM,g_{\mu\nu})$ be the  Killing development of $\left(\hyp
,{g}_{ij},{K}_{ij},{N},{Y}^{j} \right)$. It should be clear that
$(\tcM ,\widetilde{g}_{\mu\nu})$ is the universal
pseudo-Riemannian covering of $(\mcM ,g_{\mu\nu})$.

\Eqsone{rhovan}-
\eq{curvvanfu} and the Codazzi-Mainardi embedding equations
(compare \eq{ePSpcu}) show then that both $(\tcM
,\widetilde{g}_{\mu\nu})$ and $(\mcM ,{g}_{\mu\nu})$ are flat. The
remaining arguments of the proof of \cite[Theorem~4.1]{ChBeig1}
apply to show that
 $(\tcM  ,\widetilde{g}_{\mu\nu})=(\mcM,g_{\mu\nu})=(\R^{n+1},\eta_{\mu\nu})$, as desired.
\qed

\appendix
\section{Weighted H\"older spaces}
\label{definitions} \label{asymptotic}

Consider a function $f$ defined on $ \hyp_R\equiv {\R}^n\setminus
B(R) $, where $B(R)$ is a closed ball of radius $R>0$.
 We shall  write  $f = O_k(r^\beta)$ if there exists a
constant $C$ such that we have
$$
0 \leq i \leq k \qquad |\partial^i f| \leq C r^{\beta - i}.
$$
For $\sigma\in(0,1)$ we shall write
  $f = O_{k+\sigma}(r^\beta)$ if  $f = O_k(r^\beta)$ and if there exists a
constant $C$ such that we have
$$
|y-x|\le r(x)/2 \quad \Rightarrow \quad
 |\partial^k f(x) - \partial^k f(y)| \leq C |x-y|^{\sigma}
r^{\beta - k-\sigma}.
$$
Let us note that $f = O_{k+1}(r^\beta)$ implies $f =
O_{k+\sigma}(r^\beta)$ for all $\sigma\in(0,1)$, so that  the
reader unfamiliar with H\"older type spaces might wish to simply
replace, in the hypotheses of our theorems,
 the $k+\sigma$ by $k+1$ wherever
convenient.

\bigskip

\noindent
{\sc Acknowledgements} We are grateful to M. Herzlich for pointing out
an error in a previous version of the manuscript. PTC is grateful to
the Newton Institute, Cambridge, for financial support and hospitality
during the final stage of work on this paper.

\bibliographystyle{amsplain}
\bibliography{../references/hip_bib,%
../references/reffile,%
../references/newbiblio,%
../references/newbiblio2,%
../references/bibl,%
../references/howard,%
../references/bartnik,%
../references/myGR,%
../references/newbib,%
../references/Energy,%
../references/netbiblio}
\end{document}